Changes in co-publication patterns among China, the European Union (28) and the United States of America, 2016-2021

Caroline Wagner[1]*, Xiaojing Cai[2]

Abstract

The COVID-19 global pandemic starting in January 2020 disrupted international collaborations in scholarly exchange, reducing mobility and connections across the globe. An examination of Web of Science-indexed co-publications among China, the European Union (28) and the United States of America shows a drop in China-USA co-publications. Similar drops are not seen between USA and Europe or China and Europe. Moreover, the drop in China-USA cooperation can be seen beginning in 2019, before the pandemic, at a time when political tensions arose around science, technology, and innovation—with the United States claiming that China was violating intellectual property norms. The patterns suggest that political tensions, more than the pandemic, influenced the drop in China-USA cooperation.

[1] John Glenn College of Public Affairs, The Ohio State University, Columbus, OH, United States. ORCID: 0000-0002-1724-8489; wagner.911@osu.edu ; *Corresponding author
[2] Research Center for Government Governance and Public Policy, Yangzhou University, Yangzhou, Jiangsu, China. ORCID: 0000-0001-7346-6029. Email: xjcai19@gmail.com

## Introduction

International collaboration in scholarly pursuits, (including science, engineering, and social sciences, but hereafter, just 'science'), have been disrupted since January 2020 when the COVID-19 global pandemic began. Many researchers turned attention to aspects of the pandemic itself (Cai et al., 2021). In the earliest days of the pandemic, China-U.S. collaborations, measured as co-publications, increased (Fry et al., 2020). European countries were slower than China, the United States of American (USA) and the United Kingdom (UK) to publish on the pandemic. Over time, the rate of growth of China's COVID-19-related collaborations dropped, and China was less likely to be listed on international collaborative papers, in contrast to Europe, which increased engagement (Wagner et al., forthcoming). Further hindering collaborations were government-issued travel bans, visa restrictions on students and scholars from foreign nations, people wary of travel, and institutional lockdowns delaying some research projects. All these factors affected cooperative research enterprises. This paper explores the patterns of international collaboration just prior to and during the pandemic years with a focus on China, the European Union (28) and the United States.

Nations invest public funds in research and development to contribute to a knowledge-based economy, security, and health, and to enhance national prestige. Those countries whose science system is more open—inviting scholarly visitors, engaging in international collaborations—have stronger science bases (Wagner & Jonkers, 2017). A strong science base also provides nations with 'soft power' influence in international affairs. International collaborations make up about one-quarter of all collaborative articles (NSB, 2020), attracting more citations than other work and engaging elite researchers in high-profile work.

A two-decade long dramatic rise in China's investments in scientific capacity has increased its scholarly output (Leydesdorff and Zhou 2005; Zhou and Leydesdorff 2006, 2008; Zhou et al. 2009; Kostoff et al. 2007; Glänzel et al. 2008; Rousseau 2008). International collaboration has been a growing share of the rise, accounting in 2019 for 24% of China's indexed publications in Web of Science (Cao et al., 2020), and for an outsized share of citations compared to other countries (Cao et al., 2020; Wagner et al., 2015). The increases in China's scientific collaboration arises in part due to the phenomenon of Chinese student and scholar migration (Cao et al., 2020). The migrations and resulting collaborations have benefitted nations in many ways: Lee & Haupt (2021b) found that, without Chinese coauthorships, U.S. scholarly output would have dropped as a share of global output in the 2010s. Similarly, Chinese collaborations with the EU have raised numbers of publications and citations (Wang & Wang, 2017).

China's Fourteenth Five-Year Plan (Article VII, Section 3) issued by the Central Committee in 2020 supports international collaboration in science and technology (S&T). Specifics about activities are determined by companies, ministries, laboratories, hospitals, and universities (Cao et al., 2006; Serger & Breidne, 2007). China's domestic research and development spending has increased dramatically, as reported by the OECD, see Table 1 (2020). China's recent internal plans suggest that R&D spending will continue to increase, targeting a 7% per year rise from 2021-2025, with S&T and innovation projected to account for a higher percentage of GDP than the previous five years, outlined in the 13[th] Five-Year Plan period.

In addition to the disruption imposed by the global pandemic, U.S. government suspicions were raised about the depth and breadth of China's involvement in U.S. science (Lloyd-Damnjanovic & Bowe, 2020; Lee & Haupt, 2021a)**.** Political tensions rose between China and the United States around knowledge-based competition, cooperation, and scholarly collaboration. Lewis (2018) and Appelbaum et al. (2018) noted the rise in competition between the two nations, which continued into 2021. Schüller & Schüler-Zhou (2020) claimed evidence the China and USA were 'decoupling' their S&T systems. The emergence of the novel coronavirus, apparently arising in China, exacerbated political conflicts between the two nations (Devlin et al., 2020). The divisions between the two powers, and the conflicts emerging (external to science), have led some scholars to propose 'bifurcated governance' emerging around knowledge-based and wealth creating actions, such as cooperation in science or supply-chain distribution (Petricevic & Teece, 2019).

We inquire whether the recent developments have influenced China's S&T cooperation with other nations, and if so, which of the impinging factors appears to be more influential on the relationship. If the reduction in travel and study-abroad opportunities is reducing cooperation, then one would expect the lifting of travel and study-abroad restrictions could return China to global partnership. If political tensions are more influential on cooperation, then downward trends in cooperation may have longer-term consequences for global science and China's part in it. Is the collaboration between China and US affected by the global pandemic that restricts physical mobility and political tension between China and US? This article explores these questions.

## Data and Methodology

We queried Clarivate Incites for Web of Science data online to collect indicators related to research output of China, US, European Union – 28 (EU-28), and Rest of World (ROW) during 2016-2021[1]. Articles and reviews are considered; preprints are not included in the count. Full counting is used to calculate totals for internationally coauthored papers, where each involved country in a scientific publication is assigned full unit. Clarivate Web of Science--considered the 'Gold Standard' of scientific indexing services--gathers and presents data for "China Mainland" not including Hong Kong, thus, articles written by authors from China Mainland and Taiwan or Hongkong are considered as international collaborations. Although the United Kingdom left that the European Union in 2021, the EU-28 includes the United Kingdom for ease of comparison. Data were queried in January 2022 to capture all of 2021, although the database will continue to expand the 2021 collection over the next few months.

To explore the question of whether the pandemic or political tensions have had an influence on the collaborative patterns of larger economies, we first examine China, the EU-28 and the USA as the three economies producing the most scientific publications. We examine whether China, the EU-28 and the USA are as active in 2021 as before the pandemic (see also, Wagner et al., forthcoming). For reference, Table 1 shows comparative spending data from selected countries and regions as reported by the Organization for Economic Cooperation and Development (OECD), Gross Expenditures on Research and Development (GERD) in current PPP millions of US dollars between 2015 and 2019[2]. Here we see that the EU-28 spends the highest gross amount on GERD; the EU-28 has comparatively more researchers than China or the United States. As a comparison, the China-EU collaboration and USA-EU collaboration are also investigated. We compare China's cooperation with EU-28 and the United States to one another to assess whether political more than pandemic effects are influencing the relationship with China.

Table 1 Gross Expenditures on Research and Development, 2015-2019 for Selected Countries and Regions

## Results

Table 2 shows numbers of articles and reviews indexed in Web of Science from 2016 through 2021, showing the total publication output (articles and reviews) of the three largest economies and the rest of the world, with percentage shares held by each. Since 2016, China's numbers of publications have grown in number and gained higher shares each year, and the percentage share of indexed articles has steadily increased as a share of global publications. Correspondingly, the US and EU-28 show increases in numbers of articles from 2016 to 2020, with a drop in 2021 and 2021 shows a corresponding percentage share drops for these two economies. The percentage share drop for the USA and the EU-28 is taken up by China, not the rest of the world, as ROW does not gain in share. This finding continues the trend in increasing share of research publications to China identified by Leydesdorff et al. (2014).

---

[1] Data collected for each year from Clarivate Incites online, including full 2021 data as of January 28, 2022.
[2] Data are not yet available for 2020 in MSTI; data for 2015 is included as input to publications appearing in 2016.

Table 2. The number and global share of publications, China, the US, and EU-28, Articles and Reviews, full counting, 2016-2021.
Data source: Clarivate Incites online queried January 28, 2022

Figure 1 shows data from 2016-2021 but with a focus on international collaboration. International collaboration (full counts) accounts for about 25 percent of all publications—although this percentage share differs by discipline, as we will discuss. Figure 1 shows (a) the global share of international collaborations for each of the seven years studied for China, the EU-28 and the USA, as well as globally and (b) the rate at which each economy participates in international collaboration. Here one can see in Figure 1(a) the share that each economy holds of the indexed international collaborations, showing that China (Mainland) has been increasing its participation as a share of global international collaborations since 2016, although the curve starts to flatten in 2019 for reasons we discuss below. The US accounts for 40% of international collaborations worldwide, but a clear drop to 35% in the share is seen in 2021. The global share of international collaborations of EU shows a slight but steady decrease during the five years studied.

Figure 1(b) shows the rate at which economies contributes to international collaborations. Here one can see the global rate at about 25%, as mentioned above. The EU-28 is the most internationalized of the three economies—partly due to policy commitments to encourage cross-European collaborations—at close to 50% of their publications by 2021. The United States has about 40% of its publications as international collaborations. China remains the least likely to engage in international collaborations at about 25% of its publications, and dropping in 2021 to about 23%.

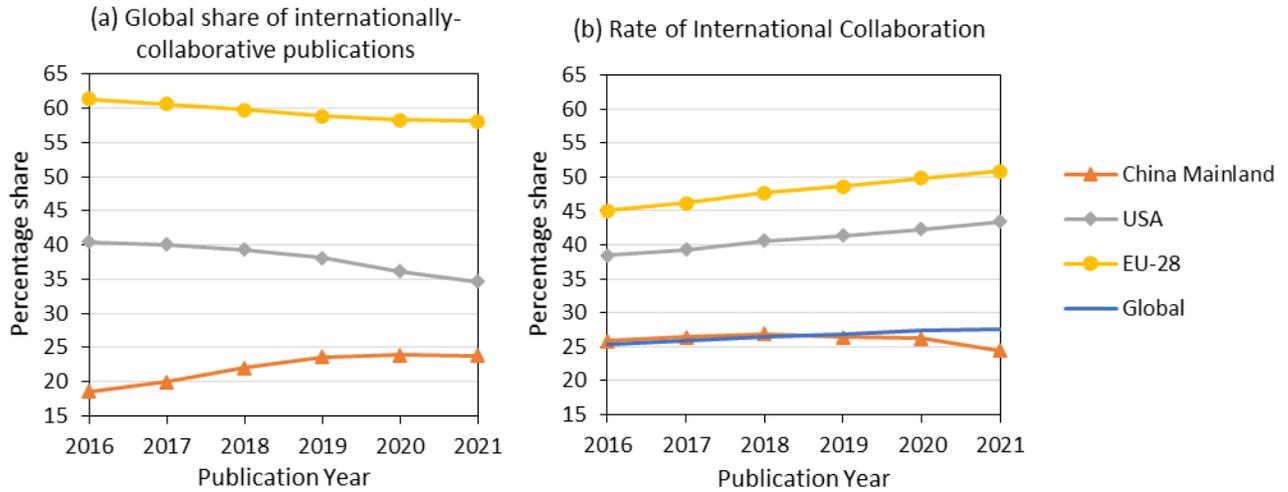

Figure 1(a) and (b). Global share of international collaborative publications (1a) and rate of international collaboration for China Mainland, the US, and EU28 (1b), 2016-2021.
Data: Clarivate Incites online, accessed January 28, 2022

The drop in China's international participation as a share of their output led us to ask where among collaborators the drop is being experienced. Figure 2 shows collaborative activity between China and the USA, between China and EU-28 and between the USA and the EU-28. It is evident that the share of China-US collaboration in global system shows a sharp decrease in 2021, from 2.71% in 2019 to 2.36% in 2021. The USA continues to collaborate more with EU-28 than with China, although that cooperation is dropping from 4.49% in 2016 to 4.17% in 2021. The collaboration between China and EU-28 continues to increase but at slowed rate in 2021.

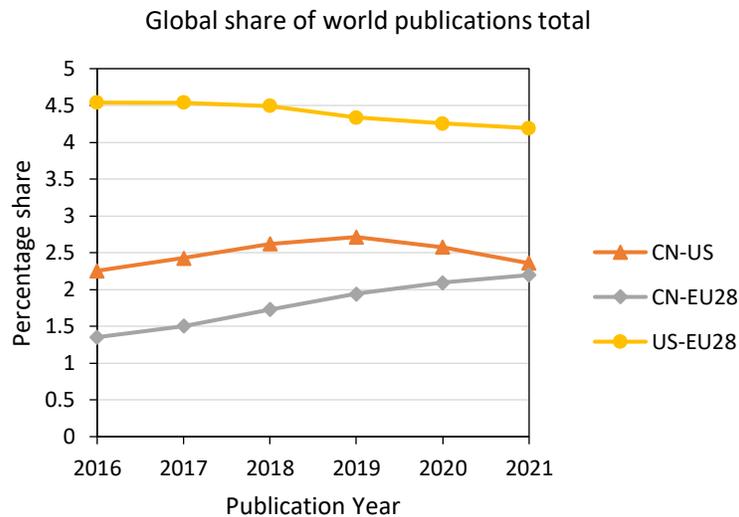

Figure 2. Percentage share of coauthorships between China-US, China-EU, and US-EU-28 in global publications, 2016-2021 (full counts)
Data: Clarivate Incites online, accessed January 28, 2022

The drop in China-USA collaborations can also be seen within the percentage share of the totals represented by the partnerships. Figure 3 shows percentage share held by EU-28 and USA collaborations dropping from about 18% of global international collaborations to 15%. China-US collaborations rose until 2019 and then drops in 2020 and 2021. China and the EU-28 rises across the period, although at a slower rate in 2021.

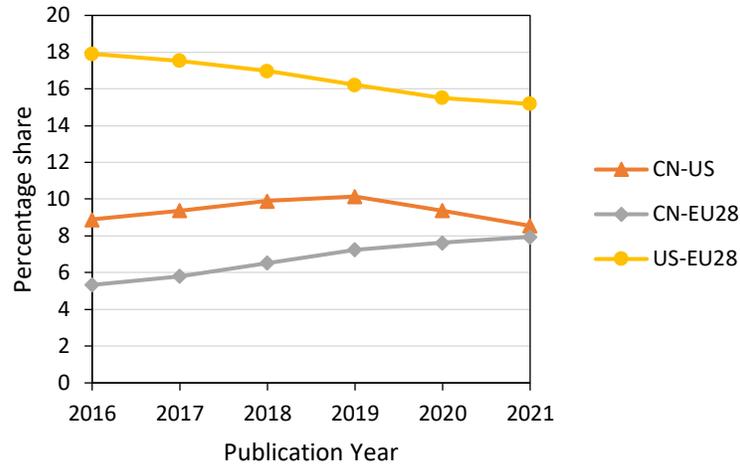

Figure 3. Percentage share of China-US, China-EU, and US-EU collaborations in global internationally-collaborated publications, 2016-2021.

Data: Clarivate Incites online, accessed January 28, 2022

We inquire further to see preferred collaborative partners for each of the large economies. Figure 4a shows the share of China's publications with the USA and the EU-28 over five years; of these shares, articles with the United States drop from 12% to 8% of all China's publications, while ones with the European Union do not drop as a share. Figure 4b shows that, around 2019, for USA publications, co-publications with China begin to drop as a share, while the share with Eu-28 grows. In other words, the USA turns more towards the EU-28 for scientific partners than to China. Figure 4c shows that, over the five years, the EU-28 is increasing co-publications with China while ones with the USA remain stable as a share of all publications.

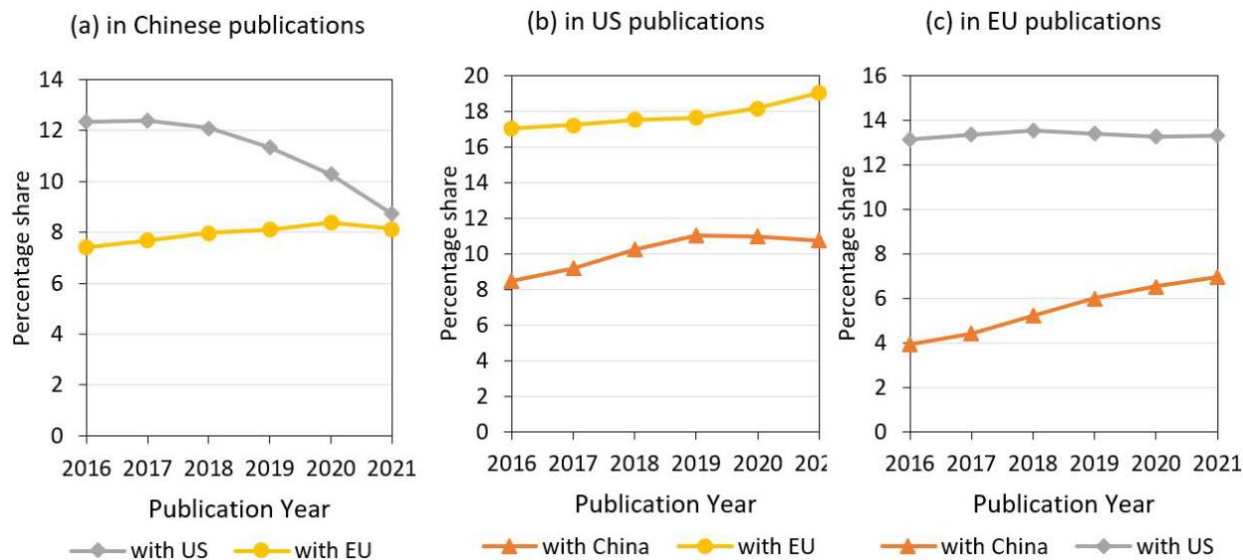

Figure 4. Percentage share of preferred partnership for China-US, China-EU, and US-EU collaborations in total publications, 2016-2021.

Data: Clarivate Incites online, accessed January 28, 2022

The drop in USA's co-publications with China around 2019 led us to explore in more detail the nature of the change. Table 3 reveals the percentage share of China-USA collaboration in Global publications by Essential Science Indicators (ESI) subjects between 2018 and 2021, when the significant changes can be noted. Table 3 shows that, in most research fields, the share of China-USA co-publication in global publications starts to decrease sharply in 2020, revealing decreasing engagement between China and the USA in most research fields.

As noted above, disciplines of science show widely different patterns of international collaborative intensity. In Table 3, one can view the shares held by China-USA collaboration by field. The fields where China's cooperation with the United States shows the most activity in percentage share are Computer Science, Geoscience, Molecular Biology, and Space Science. Of these four fields, the first three dropped between 2020 and 2021 and only Space Science shows growth. The largest drop in China-US collaborations in 2020 and 2021 are seen in Biology, Chemistry, Environment, and Microbiology. The USA-EU-28 collaborations also dropped slightly in these fields. A corresponding drop is not seen in China-EU-28 except for a small drop in Microbiology. Fields that show the biggest dropped off from 2018 for China-US cooperation are Materials Science, Microbiology, and Physics.

Increases in coauthored articles and reviews between China and the USA are seen only in Space sciences, Psychiatry/Psychology, and General Social Sciences. Increases for China with the EU-28 in 2021 are seen in Agricultural Sciences, Computer Science, Environment, Plant and Animal

Science, Psychiatry, and Social Sciences. Most fields (e.g., Biology & Biochemistry, Chemistry, Clinical Medicine, Economics & Business, Geosciences, Immunology) show little change from 2020 to 2021. For collaboration between the USA and EU-28, some research fields experience steady decline including Chemistry, Computer Science, Engineering, Environment/Ecology, Geosciences, and Mathematics. Cao et al. (2020) showed that a significant share of recorded coauthorships between China and other economies can be attributed to Chinese nationals in the partner country collaborating with someone in China. More research is needed to test whether the changes seen in Table 3 reflect underlying changes in mobility of Chinese students and researchers sojourning in foreign countries.

## Conclusion

China's international collaborations dropped sharply in 2021 overall and most notably with the United States. Multiple reasons appear to be influencing the decline. Chinese students and scholars were restricted from visiting overseas due to pandemic travel restrictions and denial of visas by the United States. Distant communications between researchers became strained by lockdowns, illness, family obligations, and funding issues. If this were simply due to the pandemic, one would expect that the same patterns would be seen for China's relationship with the EU-28 as well as with the United States, but this is not what we see.

In the late 2010s, political tensions rose between China and the United States across a number of fronts including the economy, politics, and notably around science, technology, and innovation. In 2017, the United States government opened an investigation against China under Section 301 of the Trade Act of 1974 to examine theft of American intellectual property. The US Federal Bureau of Investigation increased the number of cases investigating improper technology transfer to China. The Department of Justice reported that "80 percent of all economic espionage prosecutions brought by the U.S. Department of Justice (DOJ) allege conduct that would benefit the Chinese state, and there is at least some nexus to China in around 60 percent of all trade secret theft cases…"[3] To address the volume of reports, in 2018, the Department of Justice instituted "The China Initiative" to counter Chinese national security threats in intellectual property theft. The political tensions influenced China's cooperative S&T relationships with the United States (Lee & Haupt, 2021a). Many universities have increased restrictions on US-China collaboration activities in both countries (Silver, 2020; Lee & Haupt, 2021a).

---

[3] https://www.justice.gov/nsd/information-about-department-justice-s-china-initiative-and-compilation-china-related

High-level arrests of Chinese nationals in the United States and Chinese-American scholars also brought attention to growing tensions around intellectual property. Researchers with Chinese relations in the US were investigated (Lewis, 2021). Also, in 2018, the United States instituted visa restrictions on students and visiting scholars with additional scrutiny added to Chinese students and scholars. As early as 2019, United States universities reported a drop in the number of Chinese students coming to study. The COVID-19 pandemic, declared in January 2020, put additional and significant burdens on anyone seeking to study abroad, from the China or any other country. Visas for Chinese students were cut back to just 10 percent of what they were before the pandemic in 2020.

Restricted mobility, whether caused by political tension or pandemic, is surely having an influence on the number of international collaborations. Nevertheless, the more notable drop in number of collaborations is seen between China and the United States compared to collaborations with EU-28, and the timing beginning in 2019, rather than at the onset of the pandemic, suggests a political influence from both sides on the drop in collaboration. The majority of the drop is likely due to fewer Chinese nationals and Chinese-American scholars working with counterparts in China, although this needs more study. As shown in Table 4, the drop in number of dual-country authors in scientific publications provides direct evidence of the influence of political tensions on the research enterprise. Among all China-USA collaborative publications, the percentage share of publications with any multinational authorships or China-USA collaborations dropped significantly from 2018 to 2021, possibly indicating fewer researchers with dual positions in both countries and fewer academic exchanges.

Table 3. Multinational coauthorships in China-USA collaborative publications, 2018-2021

Data source: Clarivate Web of Science, January 20, 2022

Whether international collaborations will return to pre-pandemic levels depends on the reasons for the decline. It appears from the data that political tensions have been more influential than pandemic issues, especially for China's cooperation with the United States – which had been the largest international collaborative relationship in the world, pre-pandemic. However, political tensions of many kinds appear to have cooled the demand for visas and the offer on the US side of study and research opportunities. Individuals respond to the tensions by looking elsewhere, or looking internally, for collaborators.

It is well documented that scientists can continue to cooperate, even when their nations engage in political conflict (see Coccia, 2018). Nevertheless, recent developments have influenced China's S&T cooperation with other nations, and both political and pandemic-related factors appear to be impinging on the relationship. China's policy of civil-military fusion may be cooling interest in cooperation in the United States. To the extent that science has been enriched by the infusion of China's talent and investment in R&D, it may correspondingly be depleted by withdrawals from engagement.

## Data Availability

All data for this study will be available at figshare.

## References

Appelbaum, R. P., Cao, C., Han, X., Parker, R., & Simon, D. (2018). *Innovation in China: Challenging the global science and technology system*. John Wiley & Sons.

Cai, X., Fry, C. V., & Wagner, C. S. (2021). International collaboration during the COVID-19 crisis: autumn 2020 developments. *Scientometrics*, *126***(4)**, 3683-3692.

Cao, C., Baas, J., Wagner, C. S., & Jonkers, K. (2020). Returning scientists and the emergence of China's science system. *Science and Public Policy*, *47***(2)**, 172-183.

Cao, C., Suttmeier, R. P., & Simon, D. F. (2006). China's 15-year science and technology plan. *Physics Today*, *59***(12)**, 38.

Coccia, M. (2018). A theory of the general causes of long waves: War, general purpose technologies, and economic change. *Technological Forecasting and Social Change*, *128*, 287-295.

Devlin, K., Silver, L., & Huang, C. (2020). US views of China increasingly negative amid Coronavirus Outbreak. *Pew Research Center*, *21*.

Fry, C. V., Cai, X., Zhang, Y., & Wagner, C. S. (2020). Consolidation in a crisis: Patterns of international collaboration in early COVID-19 research. *PLoS One*, *15***(7)**, e0236307.

Glänzel, W., Debackere, K., & Meyer, M. (2008). 'Triad' or 'tetrad'? On global changes in a dynamic world. *Scientometrics*, *74***(1)**, 71-88.

Kostoff, R. N., Briggs, M. B., Rushenberg, R. L., Bowles, C. A., Icenhour, A. S., Nikodym, K. F., ... & Pecht, M. (2007). Chinese science and technology—structure and infrastructure. *Technological Forecasting and Social Change*, *74***(9)**, 1539-1573.

Lee, J., Haupt, J. (2021a) Scientific globalism during a global crisis: research collaboration and open access publications on COVID-19. *Higher Education 81***(5)**, 949-966.

Lee, J. J., & Haupt, J. P. (2021b). Scientific collaboration on COVID-19 amidst geopolitical tensions between the US and China. *The Journal of Higher Education*, *92***(2)**, 303-329.

Lewis, J.A. (2018). *Technological competition and China*. Center for Strategic and International Studies (CSIS).

Lewis, M. (2021). Time to end the U.S. Justice Department's China Initiative. Foreign Affairs Newsletter, July 22, 2021. https://foreignpolicy.com/2021/07/22/china-initiative-espionage-mistrial-hu/

Leydesdorff, L., Wagner, C. S., & Bornmann, L. (2014). The European Union, China, and the United States in the top-1% and top-10% layers of most-frequently cited publications: Competition and collaborations. *Journal of Informetrics*, *8***(3)**, 606-617.

Leydesdorff, L., & Zhou, P. (2005). Are the contributions of China and Korea upsetting the world system of science?. *Scientometrics*, *63***(3)**, 617-630.

National Science Board (2021). Publication Output: U.S. Trends and International Comparisons, https://ncses.nsf.gov/indicators

Organization for Economic Cooperation and Development, Main Science and Technology Indicators (2021). https://stats.oecd.org/Index.aspx?DataSetCode=MSTI_PUB

Petricevic, O., & Teece, D. J. 2019. The structural reshaping of globalization: Implications for strategic sectors, profiting from innovation, and the multinational enterprise. *Journal International Business Studies, 50*, 1487–1512.

Rousseau, R. (2008). Triad or Tetrad: another representation. *ISSI Newsletter*, *4***(1)**, 5-7.

Serger, S. S., & Breidne, M. (2007). China's fifteen-year plan for science and technology: An assessment. *Asia Policy*, (4), 135-164.

Schüller, M., Schüler-Zhou, Y. (2020). United States–China Decoupling: Time for European Tech Sovereignty. German Institute for Area Studies Newsletter 7, ISSN: 1862-359X.

Silver, A. (2020). Scientists in China say US government crackdown is harming collaborations. *Nature*, ***583*(7816)**, 341-343.

Wagner, C. S., Cai, X., Zhang, Y., & Fry, C. V. (forthcoming, PLoS One). One-year in: COVID-19 research at the international level in CORD-19 data. *Available at SSRN 3874974*.

Wagner, C. S., & Jonkers, K. (2017). Open countries have strong science. *Nature News*, ***550*(7674)**, 32.

Wagner, C.S., Park, H.W., Leydesdorff, L. (2015). Continuing growth of international collaboration in science: A conundrum for national governments. PLoS One, https://doi.org/10.1371/journal.pone.0131816

Wang, L., & Wang, X. (2017). Who sets up the bridge? Tracking scientific collaborations between China and the European Union. *Research Evaluation*, ***26*(2)**, 124-131.

Zhou, P., & Leydesdorff, L. (2006). The emergence of China as a leading nation in science. *Research policy*, ***35*(1)**, 83-104.

Zhou, P., & Leydesdorff, L. (2008). China ranks second in scientific publications since 2006. *ISSI Newsletter*, *4*.

Zhou P, Thijs, B., Glänzel, W. (2009). Is China also becoming a giant in the social sciences? *Scientometrics* ***71*(3)**, 593-621.

# Tables

## Table 1

| Year | USD Millions Current PPP | | | | |
|---|---|---|---|---|---|
| | 2015 | 2016 | 2017 | 2018 | 2019 |
| Selected Countries and Region | | | | | |
| Australia | 21157.08 | .. | 22376.19 | .. | 22376.19 |
| Canada | 27004.70 | 29009.68 | 29767.30 | 31268.65 | 30312.65 |
| Chile | 1552.90 | 1576.27 | 1608.73 | 1623.39 | 1623.39 |
| China (People's Republic of) | 393015.49 | 420815.58 | 465501.03 | 525693.44 | .. |
| Colombia | 2280.68 | 1799.97 | 1787.20 | 2299.83 | 2514.23 |
| Czech Republic | 6852.95 | 6369.83 | 7274.54 | 8305.38 | 8911.23 |
| European Union – 27 countries (from 01/02/2020) | 341613.94 | 360092.70 | 386684.63 | 413663.65 | 440336.57 |
| European Union – 28 countries (adding UK and EU27) | 387279.89 | 408204.04 | 437529.73 | 467897.95 | 497272.31 |
| Israel | 12666.91 | 14588.50 | 15873.75 | 17366.91 | 18740.65 |
| Japan | 168514.03 | 160269.31 | 166621.73 | 172785.86 | 173267.15 |
| Korea | 76922.04 | 80815.96 | 90289.88 | 99025.73 | 102521.44 |
| Mexico | 9577.04 | 9241.68 | 8079.11 | 7851.18 | 7407.71 |
| New Zealand | 2121.77 | .. | 2740.82 | .. | 3159.43 |
| Russia | 39012.98 | 42246.09 | 41693.19 | 44500.51 | .. |
| Singapore | 10407.37 | 10227.72 | 10530.53 | 10530.53 | .. |
| Turkey | 17734.27 | 19855.14 | 21572.18 | 23590.16 | 24243.40 |
| United Kingdom (see also EU-28 above) | 45665.95 | 48111.34 | 50845.10 | 54234.30 | 56935.75 |
| United States | 495893.00 | 522652.00 | 556343.00 | 607474.00 | 657459.00 |

Data Source: OECD MSTI January 2022
https://stats.oecd.org/Index.aspx?DataSetCode=MSTI_PUB#

## Table 2

| Year | World Total Indexed in Web of Science (only) | China Publications | Percentage share of World total | US Publications | Percentage Share of World Total | EU-28 Publications | Percentage Share of World Total | Rest of World (Excluding China, US, and EU-28) | Percentage Share of World Total |
|---|---|---|---|---|---|---|---|---|---|
| 2016 | 1,717,411 | 313,159 | 18.2 | 457,327 | 26.6 | 592,864 | 34.5 | 491,356 | 28.6 |
| 2017 | 1,790,380 | 350,298 | 19.6 | 471,850 | 26.4 | 608,466 | 34.0 | 508,282 | 28.4 |
| 2018 | 1,859,761 | 402,857 | 21.7 | 476,484 | 25.6 | 617,594 | 33.2 | 523,130 | 28.1 |
| 2019 | 2,068,920 | 494,997 | 23.9 | 509,017 | 24.6 | 669,083 | 32.3 | 575,725 | 27.8 |
| 2020 | 2,218,432 | 554,498 | 25.0 | 519,664 | 23.4 | 711,279 | 32.1 | 624,225 | 28.1 |
| 2021 | 2,153,070 | 581,017 | 27.0 | 473,943 | 22.0 | 677,942 | 31.5 | 601,040 | 27.9 |

Table 4. Global share of China-US, China-EU, US-EU collaboration in 2018-2021, by Essential Science Indicators subjects

| | China-US collaboration | | | | China-EU collaboration | | | | US-EU collaboration | | | |
|---|---|---|---|---|---|---|---|---|---|---|---|---|
| | 2018 | 2019 | 2020 | 2021 | 2018 | 2019 | 2020 | 2021 | 2018 | 2019 | 2020 | 2021 |
| Agricultural Sciences | 2.52 | 2.79 | 2.63 | 2.41 | 1.28 | 1.60 | 1.77 | 2.02 | 2.01 | 2.03 | 2.03 | 1.99 |
| Biology & Biochemistry | 3.00 | 2.92 | 2.80 | 2.43 | 1.24 | 1.45 | 1.52 | 1.54 | 5.40 | 5.42 | 5.20 | 5.38 |
| Chemistry | 2.50 | 2.51 | 2.21 | 1.75 | 1.80 | 2.02 | 2.10 | 2.10 | 2.44 | 2.34 | 2.32 | 2.17 |
| Clinical Medicine | 1.70 | 1.82 | 1.73 | 1.58 | 0.74 | 0.82 | 0.92 | 0.94 | 5.37 | 5.38 | 5.42 | 5.38 |
| Computer Science | 4.50 | 4.61 | 4.60 | 4.27 | 3.42 | 3.55 | 3.84 | 4.14 | 3.43 | 3.17 | 2.80 | 2.53 |
| Economics & Business | 2.36 | 2.64 | 2.97 | 2.95 | 1.42 | 1.66 | 2.11 | 2.25 | 6.99 | 6.79 | 6.67 | 7.08 |
| Engineering | 3.42 | 3.52 | 3.13 | 2.74 | 3.06 | 3.32 | 3.52 | 3.50 | 1.97 | 1.84 | 1.74 | 1.75 |
| Environment/Ecology | 3.51 | 3.64 | 3.18 | 2.79 | 2.46 | 2.86 | 2.88 | 2.91 | 4.83 | 4.51 | 4.05 | 3.95 |
| Geosciences | 4.82 | 5.27 | 5.01 | 4.61 | 3.69 | 4.13 | 4.52 | 4.63 | 7.19 | 6.82 | 6.81 | 6.01 |
| Immunology | 2.37 | 2.48 | 2.56 | 2.10 | 1.14 | 1.09 | 1.34 | 1.40 | 7.91 | 8.29 | 7.33 | 7.39 |
| Materials Science | 4.18 | 4.00 | 3.61 | 2.97 | 2.72 | 2.95 | 3.31 | 3.15 | 2.14 | 1.97 | 1.97 | 1.88 |
| Mathematics | 2.51 | 2.75 | 2.47 | 2.32 | 1.72 | 1.92 | 2.01 | 2.01 | 4.27 | 4.13 | 3.71 | 3.49 |
| Microbiology | 2.86 | 2.72 | 2.49 | 2.06 | 1.58 | 1.88 | 1.87 | 1.77 | 6.23 | 6.22 | 5.80 | 5.40 |
| Molecular Biology & Genetics | 4.89 | 4.45 | 4.52 | 4.02 | 1.59 | 1.85 | 1.80 | 2.08 | 8.09 | 7.56 | 7.17 | 6.95 |
| Neuroscience & Behavior | 2.46 | 2.66 | 2.46 | 2.22 | 0.95 | 1.07 | 1.22 | 1.26 | 7.92 | 8.07 | 7.89 | 7.91 |
| Pharmacology & Toxicology | 2.02 | 2.01 | 1.86 | 1.68 | 0.92 | 0.96 | 1.04 | 1.03 | 3.89 | 3.59 | 3.80 | 3.81 |
| Physics | 3.58 | 3.65 | 3.34 | 2.96 | 3.13 | 3.56 | 3.42 | 3.41 | 5.61 | 5.51 | 5.34 | 5.07 |
| Plant & Animal Science | 2.04 | 2.12 | 2.10 | 1.92 | 1.48 | 1.70 | 1.75 | 1.90 | 4.18 | 4.21 | 4.04 | 4.06 |
| Psychiatry/Psychology | 1.32 | 1.43 | 1.59 | 1.76 | 0.78 | 0.82 | 1.04 | 1.17 | 6.41 | 6.26 | 6.30 | 6.09 |
| Social Sciences, general | 1.05 | 1.18 | 1.22 | 1.34 | 0.59 | 0.70 | 0.90 | 0.99 | 3.36 | 3.26 | 3.35 | 3.48 |
| Space Science | 4.89 | 5.35 | 5.28 | 5.45 | 5.09 | 5.89 | 6.24 | 6.29 | 26.20 | 26.07 | 25.97 | 25.32 |

Note: Multidisciplinary is not shown because ESI Multidisciplinary include only multidisciplinary articles that cannot be classified into other categories based on references.

Table 4

| Year | China-USA collaborative publications total | Publications with any multinational authors | | Publications with any China-USA authors | |
|------|------|------|------|------|------|
| | | Number | Percentage share | Number | Percentage share |
| 2018 | 50,067 | 22,486 | 44.9 | 18,395 | 36.7 |
| 2019 | 55,504 | 23,877 | 43.0 | 19,069 | 34.4 |
| 2020 | 56,891 | 23,141 | 40.7 | 17,759 | 31.2 |
| 2021 | 52,714 | 19,908 | 37.8 | 14,391 | 27.3 |